\documentclass[conference]{IEEEtran}
\IEEEoverridecommandlockouts

\usepackage[utf8]{inputenc}  
\usepackage{array}           
\usepackage{graphicx}        

\usepackage{dirtytalk}       
\usepackage{fontawesome5}    

\usepackage[T1]{fontenc}
\usepackage[nocompress]{cite} 
\usepackage{microtype}
\usepackage{balance}
\usepackage{graphicx}
\usepackage{enumitem}
\usepackage{amssymb}

\usepackage{url}
\usepackage{xcolor}
\definecolor{dark-red}{rgb}{0.5,0.1,0.1}
\definecolor{dark-blue}{rgb}{0.1,0.1,0.8}
\usepackage[colorlinks,linkcolor=dark-red,citecolor=dark-blue,urlcolor=dark-blue]{hyperref}
\usepackage[nameinlink]{cleveref}

\usepackage{tikz} 
\newcommand\copyrighttext{%
  \scriptsize \textcopyright 2026 IEEE. Personal use of this material is permitted. Permission from IEEE must be obtained for all other uses, in any current or future media, including reprinting/republishing this material for advertising or promotional purposes, creating new collective works, for resale or redistribution to servers or lists, or reuse of any copyrighted component of this work in other works. Cite this article as follows: J. Vykopal, P. Čeleda, M. Horák, V. Švábenský. \textit{Technology-Enhanced Tabletop Exercises for Cybersecurity Education: Lessons Learned}. In Proceedings of the 56th IEEE Frontiers in Education Conference (FIE '26). Paphos, Cyprus, 2026. DOI: \textit{TODO add after the proceedings publication}.}
\newcommand\copyrightnotice{%
\begin{tikzpicture}[remember picture,overlay]
\node[anchor=south,yshift=5pt] at (current page.south) {\fbox{\parbox{\dimexpr\textwidth-\fboxsep-\fboxrule\relax}{\copyrighttext}}};
\end{tikzpicture}%
}

\begin{document}

\title{Technology-Enhanced Tabletop Exercises for Cybersecurity Education: Lessons Learned
\copyrightnotice
}

\author{
    \IEEEauthorblockN{\textbf{Jan Vykopal}}
    \IEEEauthorblockA{Faculty of Informatics\\
    Masaryk University\\
    Brno, Czech Republic\\
    \texttt{vykopal@fi.muni.cz}\\
    \texttt{0000-0002-3425-0951}}
\and
    \IEEEauthorblockN{\textbf{Pavel Čeleda}}
    \IEEEauthorblockA{Faculty of Informatics\\
    Masaryk University\\
    Brno, Czech Republic\\
    \texttt{celeda@fi.muni.cz}\\
    \texttt{0000-0002-3338-2856}}
\and
    \IEEEauthorblockN{\textbf{Martin Horák}}
    \IEEEauthorblockA{Faculty of Informatics\\
    Masaryk University\\
    Brno, Czech Republic\\
    \texttt{horak.martin@fi.muni.cz}\\
    \texttt{0000-0002-1835-6465}}
\and
    \IEEEauthorblockN{\textbf{Valdemar Švábenský}}
    \IEEEauthorblockA{Faculty of Informatics\\
    Masaryk University\\
    Brno, Czech Republic\\
    \texttt{valdemar@mail.muni.cz}\\
    \texttt{0000-0001-8546-280X}}
}

\maketitle


\begin{abstract}
This innovative practice full paper examines the integration of technology-enhanced tabletop exercises (TTXs) into computing education, focusing on cybersecurity curricula. The motivation is to better prepare students for complex, collaborative problem solving typical of incident response and IT governance, where coordination, communication, and timely decision-making are essential. Although TTXs are well-established in professional practice, they remain underused in universities. We address this gap by augmenting TTX delivery and evaluation through the INJECT Exercise Platform (IXP), a web-based environment that automates scenario flow and enables data-driven assessment. Our practice implements IXP to automatically deliver scenario updates, facilitate team discussions, and collect interaction data to support automated assessment. This combination enhances realism, reduces instructor workload, and provides actionable insight into student learning. From 2024 to 2026, we ran 25 exercises with 743 participants in multiple university courses and extracurricular events. We observed increased engagement and collaboration among students, and clearer visibility for instructors into how teams navigate complex scenarios. This paper shares 24 lessons learned from these exercises. Instructors and curriculum designers may benefit from concrete guidance for integrating technology-enhanced TTXs. We demonstrate that digital TTXs provide a scalable and replicable model for cybersecurity courses and others requiring team-based problem-solving.
\end{abstract}


\section{Introduction}
\label{sec:intro}

A cybersecurity tabletop exercise (TTX) is a conversation between
participants who are responsible for fulfilling a variety of roles during a cybersecurity incident~\cite{Lelewski2025cybersecurity}, such as a phishing campaign, ransomware, or denial-of-service attacks. The exercises are run by governments~\cite{NIST2006, ENISA2026methodology}, military~\cite{preda2025enhancing, ramezan2026simulating, dorton2023value}, critical infrastructure~\cite{Bartnes2017}, and private companies~\cite{osi-layer-1-2025}. The goal is to improve collaboration among roles during incident response, strengthen security awareness, reduce incident-related costs, or meet legal or contractual obligations and standards~\cite{NIS2}. 

While TTXs are widely used in professional settings, they remain underutilized in university courses~\cite{Vykopal2024research, Ottis2014, Angafor2024}. 
This paper addresses this gap by introducing a technology-enhanced approach to TTX delivery and evaluation.
The innovative practice involves implementing TTXs in cybersecurity courses using the openly available INJECT Exercise Platform (IXP)~\cite{Svabensky2024from} -- a web application designed to automate exercise delivery and enable data-driven evaluation. 

Unlike traditional pen-and-paper TTXs, IXP can automatically deliver scenario updates, facilitate team discussions, and collect interaction data for automated assessment. This approach enhances realism, reduces instructor workload, and offers actionable insights into student learning behaviors. Its uniqueness lies in combining experiential learning with analytics to improve teaching outcomes.

This paper describes the innovation and lessons learned through designing, delivering, and evaluating 25 TTXs using the platform from October 2024 to March 2026.
Automated data collection enabled analysis of team performance, communication patterns, and decision-making processes. The post-exercise feedback collected from trainees and our observations indicate improved student engagement and collaboration, and faster feedback compared to traditional teaching methods. 

This paper shares our recommendations for effective planning, development, delivery, and post-exercise reflection. We emphasize the importance of realistic, well-thought-out scenarios and of instructor support during TTX. Although our recommendations are based on experience from conducting TTXs in cybersecurity, they are not domain-specific and can be used for digital TTXs in other areas.


The paper is organized into six sections. 
\Cref{sec:related-work} introduces the TTX format and recent related work. \Cref{sec:ixp} presents the platform for the digital TTXs. \Cref{sec:process} summarizes the phases of the exercise lifecycle, which provide the structure for \Cref{sec:lessons}, where we describe the lessons learned from creating and conducting multiple TTXs using the INJECT Exercise Platform. \Cref{subsec:conclusion-materials} concludes the paper and outlines future work.

\section{Background and Related Work}
\label{sec:related-work}


A tabletop exercise is an experiential and active learning method, specifically, a form of simulation-based learning~\cite{chernikova2020simulation,Hallinger2020evolution} in small teams. Academic literature on cybersecurity TTXs and their practical delivery was reviewed by Vykopal et al. in 2024~\cite{Vykopal2024research}. The following text provides a brief overview of the TTX structure and refers readers to the review for more comprehensive information. We also summarize the most recent literature that was not included in the review.

\subsection{TTX structure}

TTXs start with a context-setting and rules briefing that explains who the exercise participants are, their roles, responsibilities, mandate, tools, and communication channels. For example, participants can be employees of a fictitious organization, working as members of a cybersecurity incident response team responsible for coordinating incident handling. They must report to the chief security officer, follow the organization's processes, and comply with national laws regulating cyber incident response. They can task other simulated participants, for example, IT administrators in the organization, when investigating an incident. The communication channel between exercise participants is a simulated email.

Once the context is set, the participants receive a number of \emph{injects}, inputs that stimulate further actions and discussions and advance the exercise~\cite{Vykopal2024research}.  
Injects can be email messages, alerts from detection systems, a new report published by a trusted organization, or breaking news in the media. Participants discuss new information within their teams and decide what to do based on their roles and the provided organizational context. For instance, if the inject is a user report on phishing received in their inbox, the participants should follow an explicitly or implicitly defined process of handling such an incident. 

Injects may not always assign concrete, clear tasks to participants. Some injects may present only a piece of information that will be useful later in the exercise, while others may task participants to create an artifact, such as a situation report for their management. Injects can also enable exploring incidents lasting for weeks in an exercise lasting only a few hours~\cite{Lelewski2025cybersecurity}.

\subsection{Recent related work}

Here, we summarize new works that were not covered in the literature review~\cite{Vykopal2024research} from 2024.
The most relevant and comprehensive is a book \emph{Cybersecurity tabletop exercises: From planning to execution}~\cite{Lelewski2025cybersecurity} that guides the reader through planning, developing, facilitating, and evaluating TTXs. It provides example scenarios for different target audiences and templates for reports from the exercise. Although the book covers tools for facilitating TTXs, it does not mention any dedicated software, only generic polling software, remote presentation, and collaboration software.


Next, four papers focus on cybersecurity TTXs.
Chowdhury and Gkioulos~\cite{Chowdhury2023} proposed a lightweight framework for conducting TTXs, evaluated through exercises involving industrial personnel and university students. The exercises were structured around the stages of the Lockheed Martin Cyber Kill Chain. They can also run fully online, leveraging common software tools and a simulator of a nuclear facility control room.
M{\"u}ller~\cite{Muller2024Ransomware} developed and evaluated a TTX for ransomware negotiations. The TTX lasted one hour and was delivered online via common software to five participants~\cite{Muller2024Ransomware-thesis}.
K{\"a}vrestad et al.~\cite{Kavrestad2025} held six TTXs lasting from 20 to 70 minutes on ransomware for a total of 90 decision-makers from various sectors. 
The participants found the TTXs valuable but too short.
Lastly, Dwight~\cite{Dwight2023collaborate} proposed a TTX format integrating the crime script analysis method used in crime investigation. However, they did not provide any evaluation.

Finally, the latest research on digital TTXs has been presented in two papers.
Watkins et al.~\cite{Watkins2026AI} introduced a platform that transforms static, scripted TTXs into dynamic, decision-responsive simulations. The platform uses AI agents and knowledge bases such as MITRE ATT\&CK to automatically generate injects. The authors presented only a proposal for the evaluation metrics (surveys, system logs) because their research has not yet been approved by their Institutional Review Board.
Sumereder et al.~\cite{Sumereder2026Digitalization} showcased digitalization of TTXs; however, for training emergency response outside cyberspace. They introduced a proof-of-concept system utilizing cameras and markers to capture physical modeling elements on a table. The approach was demonstrated using a road traffic accident scenario to show how digital tools can facilitate faster data collection, objective evaluation, and remote participation.




\section{INJECT Exercise Platform}
\label{sec:ixp}

The INJECT Exercise Platform is a lightweight, open-source web-based learning environment designed to deliver and evaluate TTXs~\cite{Svabensky2024from, inject_platform}. It transitions the traditional pen-and-paper format into a digital format that automates repetitive tasks for instructors and provides insights into student behavior. 

The platform delivers injects to teams based on time or the completion of specific \emph{milestones} to drive the scenario forward. Milestones are true/false conditions that track team progress and determine if a team has reached important situations in the scenario.

Exercise scenarios are defined in human- and machine-readable YAML files, enabling the same scenario to be deployed repeatedly under consistent conditions. The platform supports fully pre-scripted scenarios that do not require any exercise facilitators, as well as complex scenarios that evolve differently based on trainees' specific responses and inputs.

In contrast to pen-and-paper TTXs, trainees can interact with simplified versions of real-world applications, such as a simulated email client to communicate with stakeholders, a web browser to view in-exercise websites, or a firewall to block malicious network traffic.

The platform logs metadata and trainee actions, including email threads, tool usage, and reached milestones. These logs can be viewed in the built-in dashboard or exported to the JSONL format.
Instructors and exercise designers can use the dashboard to analyze team performance, such as the time required to reach milestones and the order in which they were reached. This information reveals insights into how different teams approached the same TTX scenario.

Švábenský et al.~\cite{Svabensky2024from} reported experience from using the open-source IXP to deliver a TTX in an undergraduate cybersecurity course. Three runs of the course were described. The first exercise was delivered without a specialized platform, using shared text-based documents on Microsoft SharePoint. The second run used the prototype of the IXP, and the third the very early version of the platform available in November 2023. By automating data collection and analysis, IXP offered instructors pedagogical insights into student behavior that would be laborious or practically infeasible to capture in traditional pen-and-paper TTXs.

This paper reports lessons learned from creating, delivering, and evaluating TTXs from 2024 to 2026, utilizing newer versions of the platform (v1--v5) released during this period~\cite{ixp-changelog}. IXP has evolved into a sophisticated training environment that supports diverse multimedia injects and offers flexibility for both trainees and instructors. Key advancements include support for \emph{on-demand} exercises, allowing participants to initiate sessions at their convenience, and multi-tenancy, which enables the simultaneous execution of multiple independent exercises for different groups.
Furthermore, integrated dashboards now streamline delivery and post-exercise analysis for instructors. For content development, designers can utilize a built-in exercise editor or a dedicated Visual Studio Code extension~\cite{ixp-definition} for external development or fine-tuning exercises from the editor.
 Finally, architectural optimizations now permit the platform to support up to 100 teams running in parallel.


\section{INJECT Process}
\label{sec:process}

\begin{figure}[!ht]
    \centering
    \includegraphics[width=\columnwidth]
    {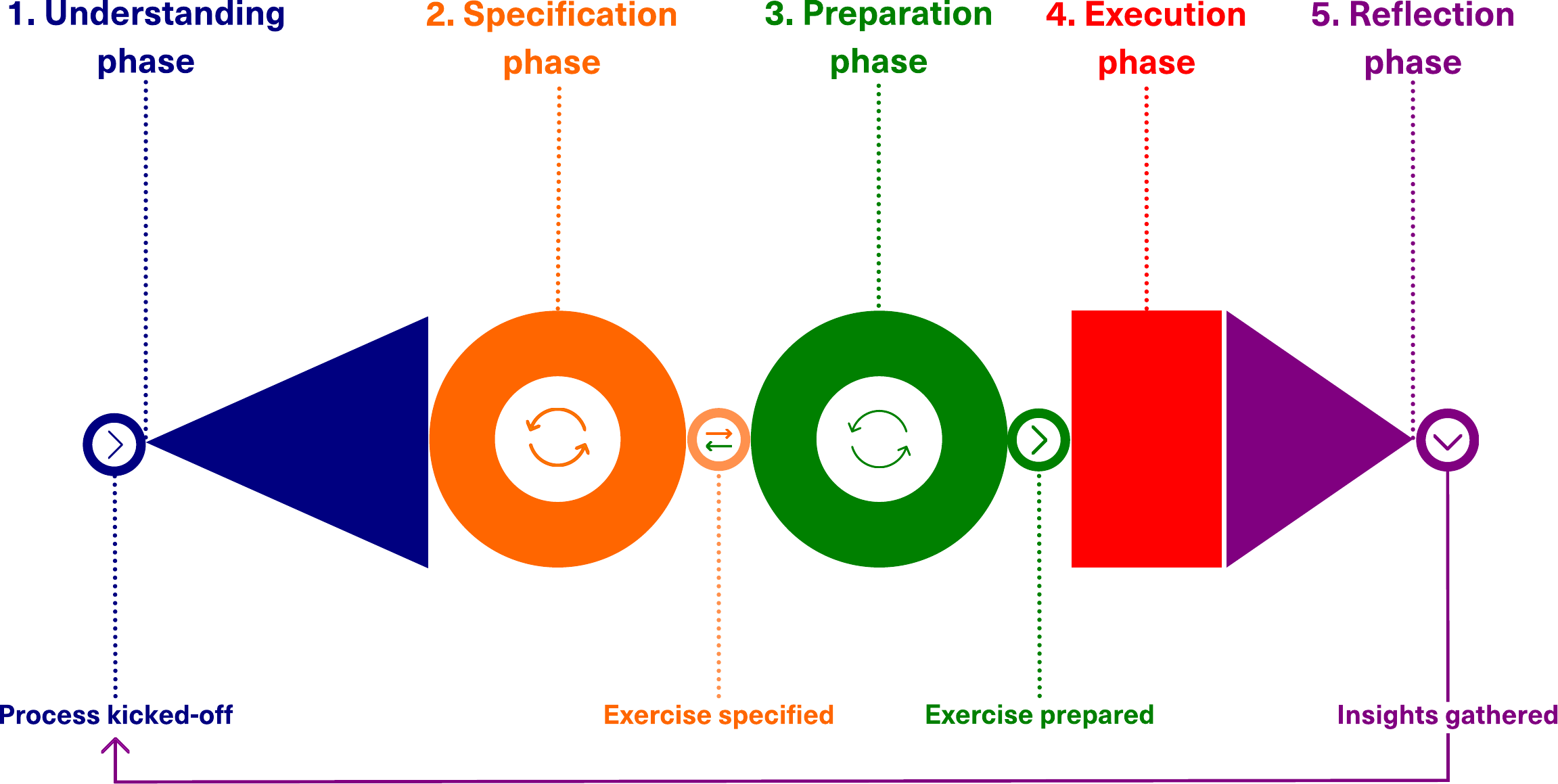} 
    \caption{INJECT Process phases~\cite{inject_process}.}
    \label{fig:process}
\end{figure}

The INJECT Process is a way to design, execute, and reflect TTXs with the IXP. The Process leverages best practices for instructional design, TTXs, and features offered by the IXP. 
It consists of five phases inspired by design thinking: \emph{understanding}, \emph{specification}, \emph{preparation}, \emph{execution}, and \emph{reflection} (\Cref{fig:process})~\cite{inject_process}.
The phases are similar to the phases of life cycles used for planning and conducting cybersecurity exercises, such as \cite{Scarfone2006guide, Vykopal2017lessons, FEMA2020HSEEP, ENISA2026methodology, Brilingaite2020framework}. However, none of these focus on TTXs in detail. The INJECT Process has been tailored for digital TTXs with respect to the IXP.

\subsection{Understanding phase}

The purpose of this phase is to establish whether the exercise is worth building at all -- and if so, for whom and toward what end. Designers and instructors are guided through methods for identifying the need behind the exercise, characterizing the target audience, and mapping the constraints that will shape subsequent design decisions. Neglecting this groundwork increases the risk that the exercise fails to achieve its intended outcomes, regardless of the quality of its technical execution. The phase also introduces the IXP's exercise types, features, and instructor involvement models, providing the knowledge required for informed decision-making in later phases.

\subsection{Specification phase}

This phase aims to produce a complete exercise specification before content is created. Separating conceptual design from implementation reduces the risk of a common failure: designers who move directly to the platform tend to oscillate between narrative decisions and technical configuration, increasing the likelihood of an incoherent final product. To design effectively at this stage, designers must connect two distinct competencies -- understanding what IXP enables, and translating learning objectives into concrete learning activities and injects. The output is a specification detailed enough that the subsequent preparation phase becomes a matter of execution.

\subsection{Preparation phase}

In this phase, exercise designers translate the exercise specification into a form that the platform can execute. Designers implement their specification either by directly editing YAML definition files or by using the platform's visual editor. This encompasses two distinct tasks: implementing the chain of events through milestone logic and conditional triggers, and creating the content -- inject texts, email templates, documents, and tool outputs -- that trainees will encounter. Before delivery, the exercise must be tested to verify that the implemented logic matches the intended design. Designers can draw on exemplary exercise definitions~\cite{ixp-scenarios} that serve as practical starting points.

\subsection{Execution phase}

The execution phase is where the experience is delivered to trainees. The exercise may run synchronously or on-demand, in person or remotely, as a one-time event or repeated across multiple cohorts. Regardless of format, the quality of execution determines whether the investment in design and preparation translates into meaningful learning. The phase covers the full arc of delivery: onboarding first-time users of the platform, providing an initial briefing that sets the context and expectations, managing the exercise as it unfolds, supporting instructors in their real-time facilitation and evaluation tasks, and concluding with a structured wrap-up (so-called “hot wash”) that transitions the exercise into reflection.

\begin{table*}[!t]
\caption{Exercises delivered from October 2024 to March 2026 through IXP version 1 to 5. Discussion exercise = \faComments. Simulation = \faProjectDiagram.}
\label{tab:exercises}
\newcommand{\simulation}{\faProjectDiagram}
\setlength{\tabcolsep}{4pt}
\begin{tabular}{lllrrl}
Type & Goal & Instructor & Exercises & Trainees & Target audience \\ \hline
\simulation & Respond to short reports as an incident response team & Yes & 2 & 113 & Cybersecurity students -- one course, two years \\
\simulation & Practice several phases of incident handling & Yes & 2 & 111 & Cybersecurity students -- one course, two years \\
\simulation & Practice writing an advisory and executive summary & Yes & 2 & 111 & Cybersecurity students -- one course, two years \\
\simulation & Practice response to a breach of personal data & Yes & 2 & 109 & Cybersecurity students -- one course, two years \\
\faComments & Train response to loss of a mobile device & No & 8 & 108 & Undergraduate students -- various disciplines \\
\faComments & Train response to loss of a mobile device & No & 3 & 45 & Students of information studies -- summer school \\
\simulation & Introduce threat modeling in maritime security & No & 2 & 56 & Cybersecurity students from two countries -- one remote \\ 
\simulation & Train handling of a phishing incident & Yes & 1 & 15 & Undergraduate computer engineering students \\ 
\simulation & Train handling of a phishing incident & Yes & 1 & 30 & 
Vocational school students and their teachers
\\ 
\simulation & Train handling of a phishing incident & Yes & 1 & 24 & Finalists of the national cybersecurity competition \\ 
\faComments & Train response to an insider threat & Yes & 1 & 21 & Finalists of the national cybersecurity competition \\ \hline
 & & & 25 & 743 &
\end{tabular}
\end{table*}

\subsection{Reflection phase}

The reflection phase closes the exercise lifecycle, but its purpose extends beyond evaluation. It operates on three levels: assessing trainee performance and connecting exercise data to actionable next steps for participants, reviewing the scenario to identify what worked and what did not, and reflecting on the overall organization and execution to improve future runs. The platform's analytics -- milestone timing, team decisions, scoring, and communication patterns -- provide a richer empirical basis for this assessment than traditional TTX formats allow. The phase yields concrete recommendations for scenario improvement, identifies facilitation strengths worth preserving, and surfaces weaknesses to address before the next delivery.

\section{Lessons learned from educational practice}
\label{sec:lessons}

Here we present lessons from 25 technology-enhanced TTXs we organized in 2024--2026 for a total of 743 trainees, mostly undergraduate students as a part of their coursework, but also high school students completing extra curricular activities.
\Cref{tab:exercises} details each exercise and its target audience.
In discussion TTXs, trainees respond to problems through questionnaires, decision tasks, and media inputs, aiming to improve communication. 
Simulations offer a realistic, process-focused experience featuring email and tool use.

The lessons were derived from structured trainee feedback systematically collected via post-exercise questionnaires and insights from instructor focus groups conducted after each event, attended by 8 instructors in total. The lessons are organized according to the INJECT Process phases (see \Cref{sec:process}).


\subsection{Understanding phase}

\paragraph{Digital format offers new possibilities}
Using a digital platform enables simulating situations that would be impractical or even impossible on paper. 
For instance, the platform can branch the scenario based on team decisions -- different choices lead to different consequences, making the exercise responsive to what trainees do rather than following a single predetermined path. Beyond this, instructors do not need to print injects and handouts for teams. The exercise can also incorporate videos or audio, or automatically and immediately request further input based on trainees’ previous actions.
This enhances the immersion and authenticity of the learning experience.
Digital TTXs have proven easy for trainees to participate in, without the need for lengthy briefings for first‑time participants before the exercise. This represents an advantage over other, more complex team-based teaching formats such as POGIL~\cite{Hu2016POGIL}.
Finally, IXP accelerates the evaluation of trainees’ inputs, which would otherwise be submitted on paper or through standard office software. When the TTX is well designed, trainees and instructors can observe and assess team performance immediately after the exercise.

\paragraph{Understand your target audience}
Creating an exercise that is valuable for \emph{everyone} is challenging. Even trainees enrolled in the same course, working in the same role, or coming from the same organization differ in their backgrounds as well as in their levels of knowledge, skills, abilities, and competencies. Narrowing the target audience can therefore simplify and focus the exercise design.
For example, if the target audience is young university students, incorporating technologies and services they are already familiar with can make the scenario more accessible and engaging.

\paragraph{The exercise format is a design decision}
The platform supports multiple exercise formats.
Choosing the most suitable format is crucial as it determines possible learning and instructor role. A simulation built around organizational processes becomes counterproductive when participants share no common background. A simulation relying on interaction with facilitators cannot be delivered to numerous trainees without a corresponding number of facilitators. A discussion-based exercise loses its value in on-demand mode, where there is no facilitator to surface disagreement or slow down premature consensus. Getting the format right requires understanding the audience and the learning objective before opening the editor.

\paragraph{Lowering the barrier to building does not eliminate the need to understand first}
IXP makes it easy to start creating. Designers can be tempted to substituting investigation of trainee needs and organizational context for personal intuition. In our experience, exercises built this way tended to cover topics the designers found interesting rather than gaps trainees faced. The result is a TTX that works technically but lands without impact. Identifying the need upfront does not slow the process down. Leveraging existing incident reports, competency frameworks, or stakeholder conversations can redirect effort toward something that will matter.

\subsection{Specification phase}

\paragraph{Every activity in the exercise should lead to learning objectives}
When specifying learning activities, designers should use the learning objectives and the target audience as a filter to decide whether an activity should be included. Having more than one designer can help, as ideas proposed by one can be challenged by another.

\paragraph{Milestone logic is the hardest cognitive shift for exercise designers}
Designers naturally think in content and narrative: what happens next, what information trainees receive, and how the story unfolds. The platform requires a different mode of thinking: states, conditions, and triggers. What action activates which milestone? What happens if the expected action never occurs? This shift from storytelling to system logic is unintuitive, and it is where specification most often breaks down. Designers who rush this step tend to produce exercises that flow well on paper but freeze in practice. Trainees miss an action the designer considered obvious, the expected milestone never activates, and the exercise stalls. Unlike traditional TTX methodologies, which have no equivalent mechanism, IXP makes this logic explicit and consequential. Getting comfortable with it during specification, before using the exercise editor, is an effective way to avoid rework later.

\paragraph{Milestones enable adapting the speed of the exercise to each team}
Milestone logic allows transforming TTXs from a simple sequence of facilitator‑driven injects delivered at predefined times into an exercise that adapts to each team's pace. For example, instead of sending an inject to all teams exactly ten minutes after the exercise begins, designers can trigger different injects based on the answers to specific questions or the use of certain tools.

\paragraph{The Tools feature enhances simulation exercises}
Simulation exercises are built around processes that typically involve actions which i) produce information essential for subsequent steps, ii) require complex or real tools or infrastructure to execute, or iii) take significantly longer to perform than the duration of the exercise itself. Compared with the traditional discussion-based exercises, a digital platform enables these actions to be simulated through in-exercise tools. Such tools automatically provide different textual or multimedia responses depending on the inputs submitted by trainees. For example, a tool can simulate a firewall capable of blocking network traffic. When a trainee enters a specific IP address, the tool activates a corresponding milestone that triggers an inject or alters the overall progress of the exercise. Most exercises we conducted simulate an incident response process, which includes several such tools. Feedback collected from trainees consistently indicates that these tools make the exercises more realistic and interactive.

\paragraph{Specification and preparation are iterative}
The INJECT Process presents the specification before preparation because a solid specification makes content creation faster and more focused. In practice, however, the boundary between the two phases is permeable. Implementation surfaces problems that were invisible at the design stage: a sequence of identical inject types that felt varied in a diagram becomes repetitive and fatiguing for trainees in the platform. A branching structure that seemed elegant on a whiteboard turns out to require more instructor intervention than possible. These discoveries are a part of the process. Specification reduces the cost of iteration, but does not eliminate it.

\paragraph{Exercise creation is difficult, even with platform support}
IXP automates delivery, structures the scenario flow, and handles data collection. This removes substantial overhead, but it does not reduce the complexity of the underlying task. Creating a good exercise requires connecting learning objectives, trainee context, narrative coherence, milestone logic, content quality, and difficulty calibration. These aspects interact: a change to the scenario flow affects the milestone logic, which affects what feedback trainees receive, which affects whether the learning objective is reached. Designers who approach the platform expecting the tooling to absorb this complexity tend to underestimate preparation time. A first exercise takes longer than anticipated -- not because the platform is hard to use, but because designing a TTX that works for a specific group of trainees is a complex task.

\subsection{Preparation phase}

\begin{figure*}[!ht]
    \centering
    \includegraphics[width=\linewidth]
    {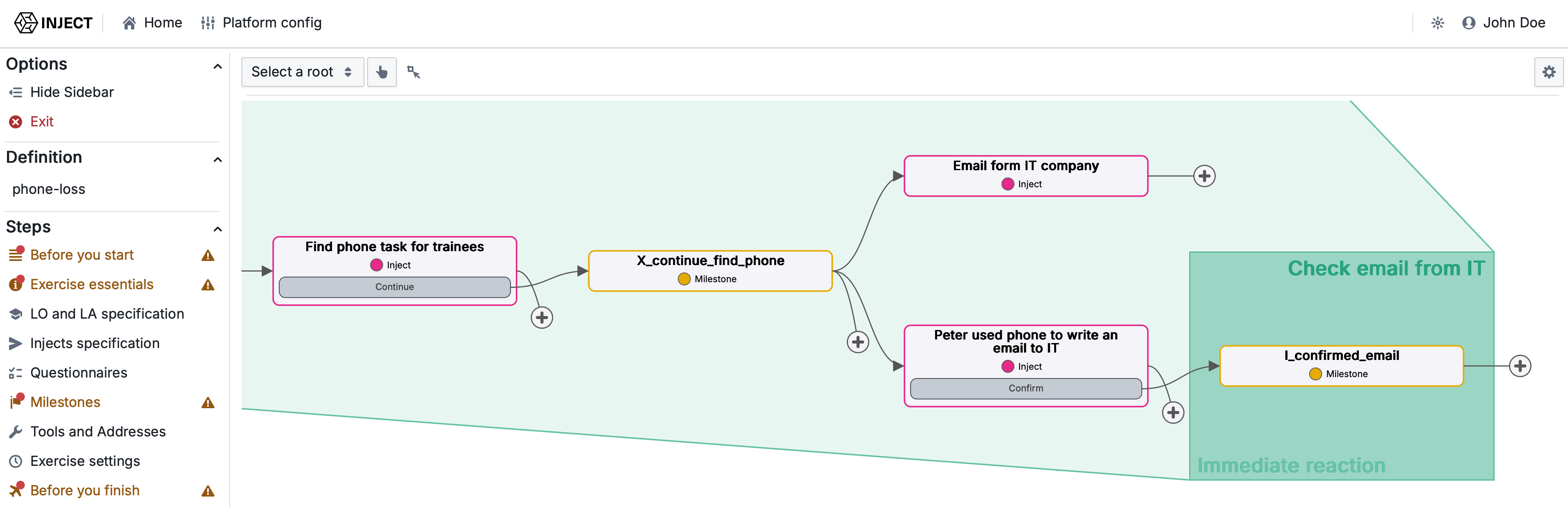} 
    \caption{Screenshot of the Editor in IXP depicting 
a part of the exercise with a learning activity with three injects connected through two milestones.}
    \label{fig:editor}
\end{figure*}

\paragraph{Digital format and new features bring new challenges}
Although the exercise definition format is human‑readable, creating multiple valid and interconnected YAML files remains challenging even for IT experts -- and nearly impossible for those without prior experience editing YAML. This barrier has been lowered by introducing two tools to IXP: an exercise editor integrated into the platform and a Visual Studio Code extension~\cite{ixp-definition}.
The editor is a web-based wizard (see \Cref{fig:editor}) that allows non‑technical users to define exercise elements and link them together into a scenario. The Visual Studio Code extension validates YAML definitions against their schema and provides code snippets for common exercise components. Designers can begin building an exercise in the editor and later fine‑tune details directly in YAML when needed.
These tools substantially simplify the work for novice designers and accelerate the workflow for experienced ones.

\paragraph{AI assistants help with specific preparation tasks but do not replace design judgment}
Generative AI tools proved useful at specific, bounded points in the preparation process: rewriting emails to match a character's voice consistently across a scenario, generating plausible variants of inject content, and identifying gaps in milestone coverage when prompted with a scenario description. What they did not replace was the judgment required to decide what the exercise should achieve, whether the milestone logic was sound, or whether the difficulty was calibrated correctly for the target audience. The useful framing is AI as a preparation accelerator for execution-level tasks -- not as a design partner for decisions that require understanding the trainees and the learning objectives.

\paragraph{Content reuse makes the model sustainable}
Creating a TTX from scratch is time-consuming. Repeated delivery is viable by reusing and adapting existing scenarios. In our experience, once a scenario is well-specified and implemented, redeploying it for a new cohort requires minimal additional effort: adjusting contextual details, refining content based on previous runs, and assigning participants to a new exercise instance. The open-source exercise library provided with the IXP~\cite{ixp-scenarios} accelerates this further by giving new designers a starting point. Sustainability of the model depends on treating scenarios as assets that accumulate value over time.

\subsection{Execution phase}

\paragraph{Two or three is a good team size} In our experience, teams of two are sometimes too small, while teams of four can be too large. Three members tend to form the optimal team size: two people may not provide enough diverse perspectives, and a third can help facilitate dialogue between members with differing views. Larger groups (more than three) often lead to reduced engagement from some participants. From a practical standpoint, forming more smaller teams may, in certain exercises, require additional instructors to manage and respond to their outputs effectively.

\paragraph{Technology lowers the facilitation bar and lends the exercise legitimacy} The platform holds the exercise structure, can deliver injects automatically, and manages the scenario flow without requiring constant instructor intervention. In practice, this means that a less experienced facilitator can run a technically sound exercise. It also means that trainees enter the experience differently -- the purpose-built environment signals that this is a structured activity. The technology itself confers credibility. The risk is that this scaffolding creates an illusion of coverage -- a facilitator can follow the tool and miss what is happening in the room. Whether teams are engaging with the scenario, whether a group is stuck or just quiet, whether a critical moment deserves to be slowed down -- these judgments remain human. The platform lowers the entry bar for execution and raises the perceived value of the exercise, but the ceiling of what it can achieve still depends on the person running it.

\begin{figure}[!t]
    \centering
    \includegraphics[width=\columnwidth]
    {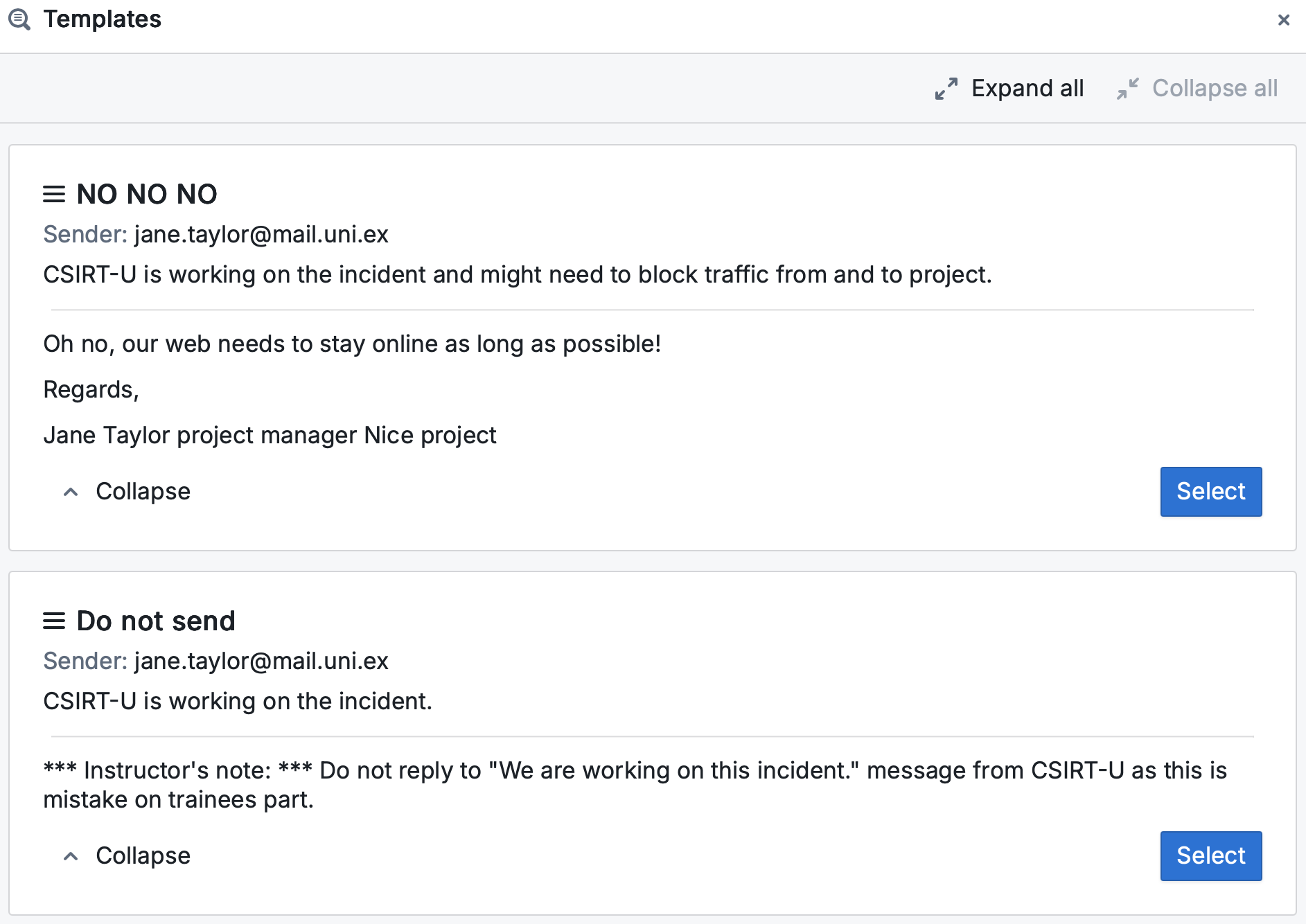} 
    \caption{Screenshot of email templates for answering trainees' mails in a TTX.}
    \label{fig:templates}
\end{figure}

\paragraph{Instructors should receive support}
If the exercise requires interactions between trainee teams and instructors, designers should use platform features that support instructors during the exercise. For example, if email communication is involved, designers need to prepare not only for the expected workflow but also for alternative responses that trainees may send. To reduce instructor workload during the exercise, designers should provide predefined email templates covering a range of possible replies to trainees (see~\Cref{fig:templates}). Another best practice is providing the instructor notes that detail actions and inputs that require instructors' assistance. This information serves as a valuable guide for novice instructors and a refresher for those who have already instructed the exercise.

\begin{figure}[!h]
    \centering
    \includegraphics[width=\linewidth]
    {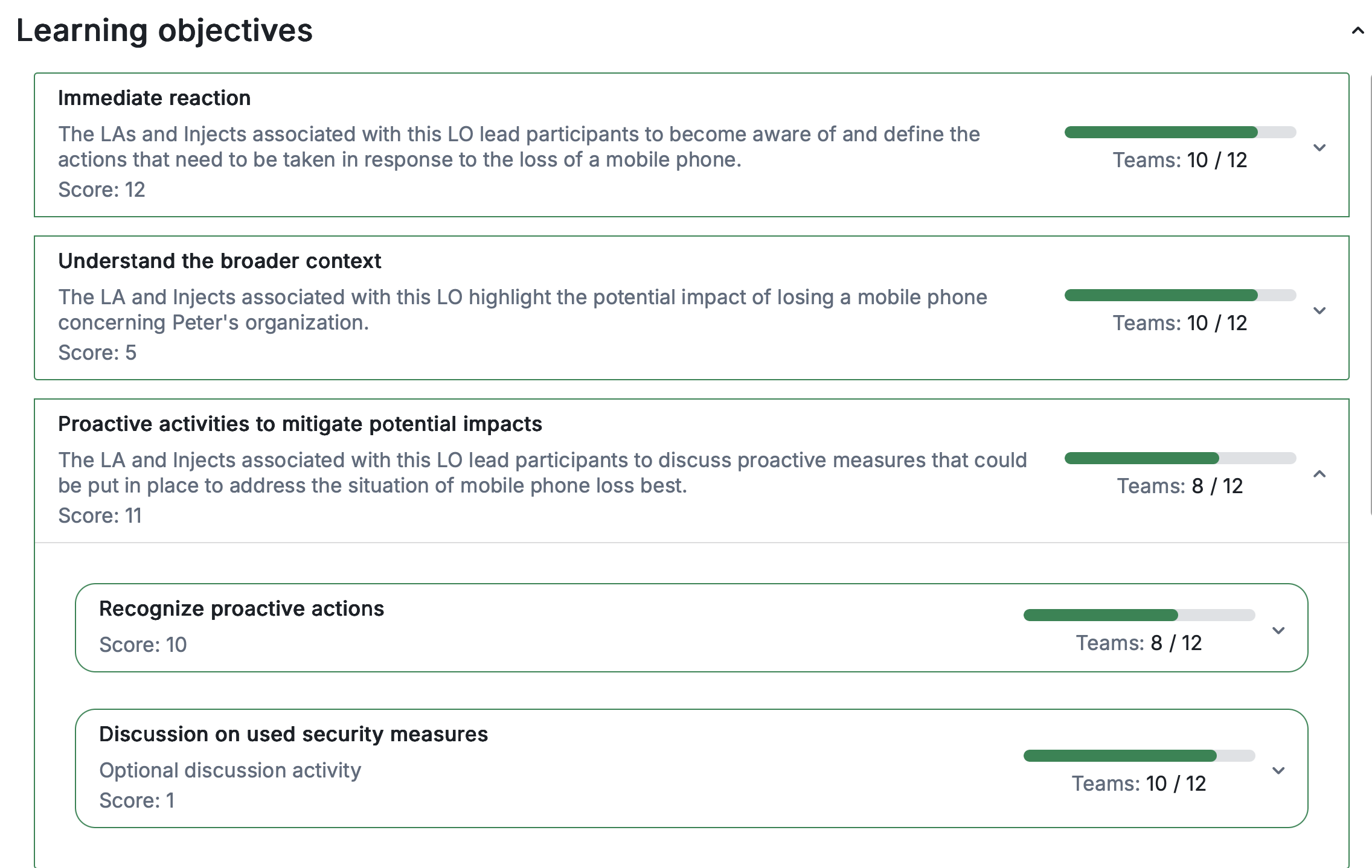} 
    \caption{Real-time view showing the achievement of learning objectives and activities by all teams in an exercise. 
    }
    \label{fig:los-las}
\end{figure}

\paragraph{Real-time view of trainees' progress is invaluable for instructors during the exercise runtime}
Instructors can immediately see the current state of the exercise for all teams. In particular, they can monitor which learning objectives have already been achieved by individual teams by completing the corresponding learning activities, each tracked through predefined milestones (see \Cref{fig:los-las}). This real‑time view provides the instructor with clear insight into whether the exercise is progressing as expected or whether any teams are struggling and may require additional support.

\paragraph{Instructor evaluation is powerful but fragile under load} The ability for instructors to evaluate trainee responses in real time is one of the platform's most distinctive features. The instructors can use predefined email templates, assess free-form submissions, or adjust milestone activation based on trainees' responses.
This enables a qualitative assessment that purely automated systems cannot replicate. The fragility appears at scale: when multiple teams reach an instructor-dependent milestone simultaneously, some teams may wait while others progress. This creates uneven pacing. One partial solution is to decouple evaluation from execution -- instructors review and score responses after the exercise concludes rather than in real time, removing the bottleneck from the critical path while preserving qualitative assessment. The tradeoff is that trainees lose immediate feedback during the exercise itself. 
A more promising direction we are currently exploring is AI-assisted evaluation: the platform can surface relevant trainee responses with suggested assessments, which instructors confirm or adjust rather than generating from scratch. Early experience suggests this substantially reduces evaluation time without sacrificing the human judgment that makes the assessment meaningful.

\paragraph{Remote exercise runs should be accompanied by an external communication channel}
We recommend setting up a synchronous communication channel outside the exercise platform, such as a video conference. The channel can be used for pre‑ and post‑exercise briefings, as well as virtual breakout rooms to support fully distributed teams.

\paragraph{Fully automated exercises scale best but might lack flexibility}
The platform makes it possible to deliver exercises that each team can start independently, at a time that suits them. Unlike synchronous exercises, which begin simultaneously for all teams, these on‑demand exercises cannot rely on human facilitators, as they are not continuously available. Consequently, on‑demand exercises must be fully automated. While this approach allows the exercise to be offered to hundreds of participants, it also restricts the design to exercise elements that can be automatically evaluated, such as multiple‑choice questionnaires. These constraints may lead to less flexible scenarios, potentially reducing the trainees’ overall experience.

\paragraph{Once the exercise is digital, trainees compare it to other digital experiences -- not to TTX standards} A paper-based TTX is evaluated against other paper-based TTXs. A digital one is evaluated against every other digital experience trainees have encountered -- polished applications, responsive interfaces, seamless interactions. This is not an unreasonable expectation, but it is different, and it catches designers off guard. Minor friction that would be invisible in a traditional format -- a slow page load, an unintuitive navigation step, a visual that does not render cleanly -- becomes noticeable and affects perceived quality. The implication is not that the platform needs to compete with consumer software, but that expectation management and interface quality carry more weight in digital delivery than experience with traditional TTXs would suggest.

\begin{figure}[!t]
    \centering
    \includegraphics[width=0.95\columnwidth]
    {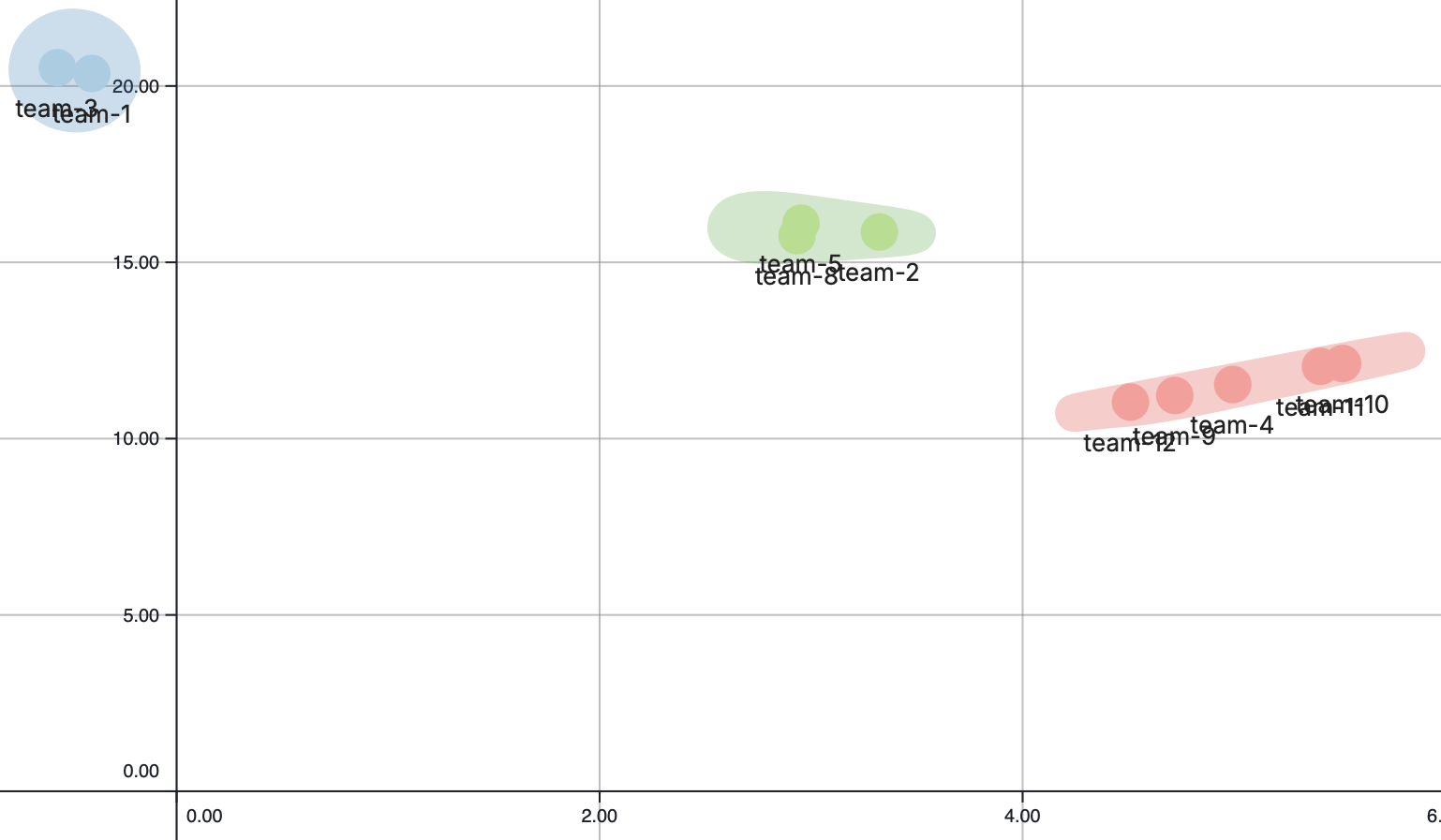}
    \caption{
    Analyst view showing team milestones grouped into three clusters (blue, green, and red) by activity similarity. Since the axes represent abstract dimensions, the key factor is the distance between the clusters, which guides the analyst's focus.
    }
    \label{fig:clusters}
\end{figure}

\begin{figure*}[!t]
    \centering
    \includegraphics[width=\linewidth]
    {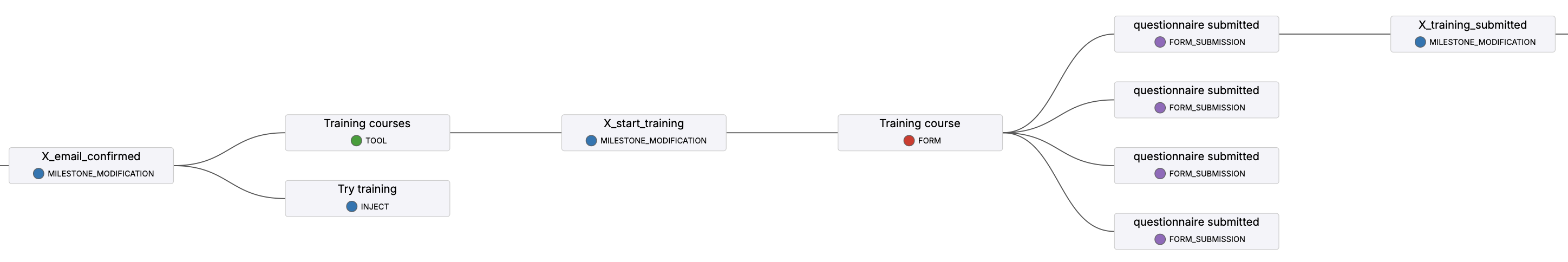} 
    \caption{
    Analyst view showing a graph of selected exercise events for a team. Activating the left milestone triggered an inject and tool access. Tool use then activated another milestone that opens a form. The team made four submissions, and one of them activated the milestone on the right. For example, this view can be helpful for revealing unexpected progression that may not be intended by the scenario designer.}
    \label{fig:cause-effect}
\end{figure*}

\subsection{Reflection phase}

\paragraph{The digital format enables detailed post-exercise analyses for improving the exercise}
All trainees' inputs and actions are logged and processed by the platform, which provides valuable data and insights not only for reflection but also for improvement of the exercise itself. This is particularly useful after the first run of a new or updated exercise. We use the provided data to answer the following design questions: 
\begin{itemize}
    \item \emph{Did the majority of teams achieve all learning objectives or high score} (see \Cref{fig:los-las})? If not, the exercise might not match their proficiency and needs to be revised.
    \item \emph{Did the teams follow the anticipated progression, or were there significant deviations?} \Cref{fig:clusters} illustrates the analytical results addressing this question. Do these anomalies represent valid, but unexpected solutions aligned with the learning objectives, or do they expose weaknesses in the scenario design that require revision? The answer to this question can be found in the causal graph of exercise events, see \Cref{fig:cause-effect}.
    \item \emph{What is the qualitative assessment from the participants?} We conclude the exercise by asking trainees for feedback. Having their answers in the platform allows for more convenient analysis of their experience compared to the feedback collected via external survey tools.
\end{itemize}



Furthermore, the ability to export exercise logs in JSONL format is invaluable for external data processing. This feature streamlines computing education research, which can, in turn, influence the practical implementation of digital TTXs.

\paragraph{Multiple exercises for the same group of trainees bring benefits}
Once trainees receive their accounts in the platform, they can be easily assigned to more exercises without additional overhead of creating new accounts. What is more, instructors can see and compare trainee performance across exercises.

\paragraph{The exercise is the trigger -- reflection is where the learning happens}
Trainees engage with the scenario during execution but the deeper processing comes afterward. This includes understanding why their decisions were right or wrong, connecting the exercise to their organizational context, and identifying what they would do differently. All of this happens in the conversation that follows. In our experience, trainees consistently wanted more time for debriefing than the schedule allocated. They wanted to compare their decisions with other teams, hear the reasoning behind the scenario design, and discuss whether their responses would hold in reality. 
IXP supports this: milestone timing, team decisions, scoring, and communication patterns are all captured and available for structured review in ways that paper-based TTXs cannot match. The risk is that this data goes unused. Trainees leave with an experience rather than anything concrete to act on -- no checklist or commitment to a specific next step. A structured debriefing that connects TTX data to actionable outputs would substantially increase the probability of behavioral change of trainees after the exercise. 
This value, however, depends on a facilitator who can surface disagreement, slow down premature consensus, and anchor the discussion in what happened. On-demand exercises sacrifice this. They are the most scalable format, but scaling delivery at the cost of structured reflection trades the exercise's most durable benefit for its most convenient form. Treating reflection as a procedural closing step, or removing it entirely, leaves significant potential unrealized.

\section{Conclusions and Future Work}
\label{subsec:conclusion-materials}

TTXs are widely used in professional practice but remain relatively uncommon in higher education. Based on our experience conducting 25 exercises with 743 participants, 
we demonstrated that TTXs can be effectively integrated into university-level teaching when supported by a dedicated platform. We presented 24 lessons learned from two years of designing, developing, running, and evaluating these exercises using the open-source INJECT Exercise Platform~\cite{inject_platform}.

The lessons share a common theme: the value of technology-enhanced TTXs is not determined solely by the platform. IXP removes significant problems: manual delivery, paper-based data collection, and coordination overhead. However, it does not remove the complexity of designing TTXs that achieve learning outcomes, facilitating experiences that trigger reflection, or establishing the organizational conditions in which exercises lead to behavioral change. In each phase of the lifecycle, we found that the platform's potential was either realized or diminished based on decisions and practices outside the tool.
Ultimately, these insights underscore that the \emph{human-in-the-loop} remains an indispensable element in the exercise life cycle. Our objective is therefore not to replace the educator, but to provide a robust framework that supports and augments their expertise. In this view, technology serves as an enabler of, rather than a substitute for, high-quality human instruction.

The lessons are relevant to educators integrating digital TTXs 
and developers of similar platforms. These insights highlight where developers can focus their efforts to address current limitations and enhance support for instructors and learners.

\subsection{Future work}

Because exercise specification and preparation are complex, demanding, and time‑consuming phases, there is potential for using language models to generate TTX scenarios or their components, such as individual injects. Language models may also support the analysis of trainees’ textual responses during and after an exercise, as well as the generation of reports for trainees and instructors. While large language models typically achieve higher performance, smaller models may be more suitable in cybersecurity education contexts, as they can be run locally without sharing sensitive data with third parties.

Another direction for future work is enhancing post-exercise reflection. Both our experience and existing research~\cite{Tembrevilla2024experiential,Rover2021reflective} emphasize that reflection is a crucial phase. Yet it remains insufficiently addressed, as most attention is devoted to earlier phases: planning, preparation, and execution. Improving structured reflection processes and tooling could strengthen learning outcomes and support long-term skill development.

\section*{Acknowledgment}
This research was supported by the Open Calls for Security Research 2023--2029 (OPSEC) program granted by the Ministry of the Interior of the Czech Republic under No. VK01030007 -- \textit{Intelligent Tools for Planning, Conducting, and Evaluating Tabletop Exercises}.

\bibliographystyle{IEEEtran}
\bibliography{references.bib}

@misc{NIS2,
  author       = {{European Parliament} and {Council of the European Union}},
  title        = {Directive ({EU}) 2022/2555 of the {E}uropean {P}arliament and of the {C}ouncil of 14 {D}ecember 2022 on measures for a high common level of cybersecurity across the {U}nion},
  year         = {2022},
  howpublished = {Official Journal of the European Union, L 333/80},
  url          = {http://data.europa.eu/eli/dir/2022/2555/oj},
  note = {},
  note         = {Amending Regulation (EU) No 910/2014 and Directive (EU) 2018/1972, and repealing Directive (EU) 2016/1148 (NIS Directive)}
}

@article{Tembrevilla2024experiential,
author = {Tembrevilla, Gerald and Phillion, André and Zeadin, Melec},
title = {Experiential learning in engineering education: A systematic literature review},
journal = {Journal of Engineering Education},
volume = {113},
number = {1},
pages = {195-218},
doi = {https://doi.org/10.1002/jee.20575},
url = {https://doi.org/10.1002/jee.20575},
eprint = {https://onlinelibrary.wiley.com/doi/pdf/10.1002/jee.20575},
year = {2024}
}

@INPROCEEDINGS{Rover2021reflective,
  author={Rover, Diane T. and Duwe, Henry J. and Mina, Mani and Fila, Nicholas D. and Jones, Phillip H. and Sleeth, Lindsey S.},
  booktitle={2021 IEEE Frontiers in Education Conference (FIE)}, 
  title={{Learning and Professional Development Through Integrated Reflective Activities in Electrical and Computer Engineering Courses}}, 
  year={2021},
  volume={},
  number={},
  pages={1-9},
  doi={10.1109/FIE49875.2021.9637478},
  url={https://doi.org/10.1109/FIE49875.2021.9637478}
}

@inproceedings{Hu2016POGIL,
author = {Hu, Helen H. and Kussmaul, Clifton and Knaeble, Brian and Mayfield, Chris and Yadav, Aman},
title = {{Results from a Survey of Faculty Adoption of Process Oriented Guided Inquiry Learning (POGIL) in Computer Science}},
year = {2016},
isbn = {9781450342315},
publisher = {Association for Computing Machinery},
address = {New York, NY, USA},
url = {https://doi.org/10.1145/2899415.2899471},
doi = {10.1145/2899415.2899471},
booktitle = {Proceedings of the 2016 ACM Conference on Innovation and Technology in Computer Science Education},
pages = {186–191},
numpages = {6},
location = {Arequipa, Peru},
series = {ITiCSE '16}
}

@misc{ixp-scenarios,
    author = {{INJECT Team}},
    title = {{Available Exercise Definitions}},
    howpublished = {\url{https://docs.inject.muni.cz/INJECT_process/available-definitions/}},
    year = {2026},
    note = {Accessed: March 11, 2026}
}

@misc{ixp-changelog,
    author = {{INJECT Team}},
    title = {{Platform Changelog}},
    howpublished = {\url{https://docs.inject.muni.cz/changelog/}},
    year = {2026},
    note = {Accessed: March 11, 2026}
}

@misc{ixp-definition,
    author = {{INJECT Team}},
    title = {{IXP-Definition -- MS Visual Studio Code Extension}},
    howpublished = {\url{https://marketplace.visualstudio.com/items?itemName=inject-muni.ixp-definition}},
    year = {2026},
    note = {Accessed: March 11, 2026}
}

@misc{Scarfone2006guide,
  author       = {Scarfone, Karen and Grance, Tim and Sexton, Riley},
  title        = {{NIST Special Publication 800-84: Guide to Test, Training, and Exercise Programs for IT Plans and Capabilities}},
  howpublished = {\url{https://csrc.nist.gov/pubs/sp/800/84/final}},
  year         = {2006},
  month        = {Sep},
  note         = {Accessed: March 11, 2026}
}

@article{Brilingaite2020framework,
title = {A framework for competence development and assessment in hybrid cybersecurity exercises},
journal = {Computers \& Security},
volume = {88},
pages = {101607},
year = {2020},
issn = {0167-4048},
doi = {https://doi.org/10.1016/j.cose.2019.101607},
url = {https://doi.org/10.1016/j.cose.2019.101607},
author = {Agnė Brilingaitė and Linas Bukauskas and Aušrius Juozapavičius}
}

@misc{FEMA2020HSEEP,
  author       = {{Federal Emergency Management Agency}},
  title        = {{Homeland Security Exercise and Evaluation Program (HSEEP)}},
  howpublished = {\url{https://preptoolkit.fema.gov/documents/1269813/1269861/HSEEP_Revision_Jan20_Final.pdf}},
  year         = {2020},
  month        = {jan},
  note         = {Accessed: March 11, 2026}
}

@techreport{ENISA2026methodology,
  author      = {Zacharis, Alexandros and Sarri, Anna and Van Heurck, Christian and Fanourakis, Fanouris and Fernández, Gema and Christoforatos, Nikolaos and Arcus, Radu},
  title       = {{The ENISA Cybersecurity Exercise Methodology}},  title     = {{The ENISA Cybersecurity Exercise Methodology: End-to-end guide on how to plan, run and evaluate an exercise}},
  institution = {European Union Agency for Cybersecurity (ENISA)},
  year        = {2026},
  month       = {2},
  version     = {1.0},
  type        = {Technical Report},
  isbn        = {978-92-9204-735-1},
  doi         = {0.2824/4949728},
  url         = {https://www.enisa.europa.eu/sites/default/files/2026-02/The%20ENISA%20Cybersecurity%20Exercise%20Methodology.pdf}
}

@INPROCEEDINGS{Vykopal2017lessons,  
    author={J. Vykopal and Vizvary, Martin and Oslejsek, Radek and Celeda, Pavel and Tovarnak, Daniel},  
    booktitle={2017 IEEE Frontiers in Education Conference (FIE)},   
    title={Lessons Learned From Complex Hands-on Defence Exercises in a Cyber Range},   
    year={2017},  
    volume={},  
    number={},  
    pages={1-8},  
    isbn={978-1-5090-5920-1},
    doi={10.1109/FIE.2017.8190713},
    url={https://doi.org/10.1109/FIE.2017.8190713}
}

@misc{inject_process,
  author = {{INJECT Team}},
  title = {{INJECT Process}},
  howpublished = {\url{https://docs.inject.muni.cz/INJECT_process/intro/overview/}},
  year = {2026},
  note = {{Documentation of designing tabletop security exercises with the INJECT Exercise Platform, accessed: March 20, 2026}}
}

@misc{inject_platform,
  author = {{INJECT Team}},
  title = {{INJECT Exercise Platform}},
  howpublished = {\url{https://inject.muni.cz}},
  year = {2026},
  note = {{Open-source platform for tabletop exercises, accessed: March 20, 2026}},
}

@InProceedings{Sumereder2026Digitalization,
author="Sumereder, Anna
and B{\"u}rger, Bernhard
and Woitsch, Robert",
editor="Seb{\"o}ck, Walter
and Lampoltshammer, Thomas J.
and Dugdale, Julie
and Zeller, Ingeborg",
title="{Digitalization of Table-Top Exercises: An Emergency Response Training Showcase}",
booktitle="Information Technology in Disaster Risk Reduction",
year="2026",
publisher="Springer Nature Switzerland",
address="Cham",
pages="49--65",
url={https://doi.org/10.1007/978-3-031-97115-0_4},
isbn="978-3-031-97115-0"
}

@INPROCEEDINGS{Watkins2026AI,
  author={Watkins, Trenton and Davis, Ben and Ponnuru, Raveendra Babu and Azab, Mohammed},
  booktitle={2026 IEEE 16th Annual Computing and Communication Workshop and Conference}, 
  title={{AI-Driven Immersive Emulation for Tabletop Scenarios}}, 
  year={2026},
  pages={234-240},
  doi={10.1109/CCWC67433.2026.11393711}
}

@inproceedings{Dwight2023collaborate,
  title={Collaborate, design, and generate cybercrime script tabletop exercises for cybersecurity education},
  author={Dwight, Joshua},
  booktitle={International Conference on Computers in Education},
  year={2023},
  url = {https://library.apsce.net/index.php/ICCE/article/view/1406/1300}
}

@misc{Muller2024Ransomware-thesis,
  author       = "M{\"u}ller, Lea",
  title        = {Tabletop Exercise for Ransomware Negotiations},
  school       = {Albstadt-Sigmaringen University, Faculty of Computer Science},
  year         = {2024},
  address      = {Germany},
  month        = {February},
  type         = {Bachelor's Thesis},
  note         = {{Bachelor's thesis}},
  url = {https://www.researchgate.net/publication/381290646_Tabletop_Exercise_for_Ransomware_Negotiations}
}

@InProceedings{Muller2024Ransomware,
author="M{\"u}ller, Lea",
editor="Schmorrow, Dylan D.
and Fidopiastis, Cali M.",
title="{Tabletop Exercise for Ransomware Negotiations}",
booktitle="Augmented Cognition",
year="2024",
publisher="Springer Nature Switzerland",
address="Cham",
pages="166--184",
url={https://doi.org/10.1007/978-3-031-61572-6_12},
isbn="978-3-031-61572-6"
}

@InProceedings{Chowdhury2023,
author="Chowdhury, Nabin and Gkioulos, Vasileios",
title="{A Framework for Developing Tabletop Cybersecurity Exercises}",
booktitle="Computer Security. ESORICS 2022 International Workshops",
year="2023",
publisher="Springer International Publishing",
address="Cham",
pages="116--133",
url={https://doi.org/10.1007/978-3-031-25460-4_7},
isbn="978-3-031-25460-4"
}

@article{Hallinger2020evolution,
author = {Philip Hallinger and Ray Wang},
title ={{The Evolution of Simulation-Based Learning Across the Disciplines, 1965–2018: A Science Map of the Literature}},
journal = {Simulation \& Gaming},
volume = {51},
number = {1},
pages = {9-32},
year = {2020},
doi = {10.1177/1046878119888246},
URL = {https://doi.org/10.1177/1046878119888246},
eprint = {https://doi.org/10.1177/1046878119888246}
}

@article{ramezan2026simulating,
  title={Simulating Cyber-Resilience: The Strategic Role of the Locked Shields Exercise in Enhancing International Cyber Preparedness},
  author={Ramezan, Christopher A and Schaupp, Ludwig Christian and Vitullo, Elizabeth A and Walker, William J},
  journal={Journal of Cybersecurity Education, Research and Practice},
  volume={2026},
  number={1},
  pages={5},
  year={2026},
  url={https://doi.org/10.62915/2472-2707.1260}
}

@article{preda2025enhancing,
  title={Enhancing Civil-Military Cyber Resilience Lessons from the ECYBRIDGE Tabletop Exercise.},
  author={Preda, Marius and Popescu, Valeria and Argint, Cornel and Iancu, Niculae and Raicu, Gabriel and Ene, Gabriel},
  journal={International Journal of Information Security \& Cybercrime},
  volume={14},
  number={1},
  year={2025},
  url={https://www.ceeol.com/search/article-detail?id=1349715}
}

@article{Angafor2024,
title = {{MalAware: A tabletop exercise for malware security awareness education and incident response training}},
journal = {Internet of Things and Cyber-Physical Systems},
volume = {4},
pages = {280-292},
year = {2024},
issn = {2667-3452},
doi = {https://doi.org/10.1016/j.iotcps.2024.02.003},
url = {https://doi.org/10.1016/j.iotcps.2024.02.003},
author = {Giddeon Angafor and Iryna Yevseyeva and Leandros Maglaras}
}

@article{Ottis2014,
   author = {Ottis, Rain},
   doi = {10.1515/jhsem-2014-0031},
   issn = {1547-7355},
   issue = {4},
   journal = {Journal of Homeland Security and Emergency Management},
   month = {12},
   pages = {579-592},
   title = {{Light Weight Tabletop Exercise for Cybersecurity Education}},
   volume = {11},
   url = {https://doi.org/10.1515/jhsem-2014-0031},
   year = {2014},
}

@article{Bartnes2017,
title = {{Challenges in IT security preparedness exercises: A case study}},
journal = {Computers \& Security},
volume = {67},
pages = {280-290},
year = {2017},
issn = {0167-4048},
doi = {https://doi.org/10.1016/j.cose.2016.11.017},
url = {https://doi.org/10.1016/j.cose.2016.11.017},
author = {Maria Bartnes and Nils Brede Moe}
}

@misc{osi-layer-1-2025,
  title={{Cyber-Security at OSI Layer 1: Defence-in-Depth for Energy Grids}},
  author={Lenk, Peter J and Kines, Stephen and Taube, Erki and M{\"u}nzer, Markus and Kuusk, Tanel and Alberico, Stefano},
  note = {{MP-SAS-190 Technical Evaluation Report}},
  year = {2025},
  url = {https://publications.sto.nato.int/publications/STO%20Meeting%20Proceedings/STO-MP-SAS-190/MP-SAS-190-07.pdf}
}

@inproceedings{dorton2023value,
  title={{The Value of Wargames and Tabletop Exercises as Naturalistic Tools}},
  author={Dorton, Stephen L and Fersch, Theresa and Barrett, Emily and Langone, Andrew and Seip, Mark and Bilsborough, Shane and Hudson Jr, Curtis B and Ward, Paul and Neville, Kelly J},
  booktitle={Proceedings of the Human Factors and Ergonomics Society Annual Meeting},
  volume={67},
  number={1},
  pages={2454--2459},
  year={2023},
  organization={SAGE Publications Sage CA: Los Angeles, CA},
  url={https://doi.org/10.1177/21695067231192617}
}

@TechReport{NIST2006,
    author      = {Tim Grance and Tamara Nolan and Kristin Burke and Rich Dudley and Gregory White and Travis Good},
    title       = {Guide to Test, Training, and Exercise Programs for IT Plans and Capabilities},
    institution = {NIST},
    year        = {2006},
    month       = {09},
    doi         = {10.6028/NIST.SP.800-84}
}

@book{Lelewski2025cybersecurity,
  title={{Cybersecurity Tabletop Exercises: From Planning to Execution}},
  author={Lelewski, Robert and Hollenberger, John},
  year={2025},
  publisher={No Starch Press},
  address={San Francisco, USA},
  isbn={1718503822},
  note       = {{ISBN: 978-1718503823}}
}

@inproceedings{Svabensky2024from,
    author    = {\v{S}v\'{a}bensk\'{y}, Valdemar and Vykopal, Jan and Hor\'{a}k, Martin and Hofbauer, Martin and \v{C}eleda, Pavel},
    title     = {{From Paper to Platform: Evolution of a Novel Learning Environment for Tabletop Exercises}},
    booktitle = {Innovation and Technology in Computer Science Education},
    publisher = {ACM},
    address   = {New York, NY, USA},
    year      = {2024},
    pages     = {213--219},
    numpages  = {7},
    isbn      = {979-8-4007-0600-4},
    url       = {https://doi.org/10.1145/3649217.3653639},
    doi       = {10.1145/3649217.3653639},
}

@inproceedings{Vykopal2024research,
    author    = {Vykopal, Jan and \v{C}eleda, Pavel and \v{S}v\'{a}bensk\'{y}, Valdemar and Hofbauer, Martin and Hor\'{a}k, Martin},
    title     = {{Research and Practice of Delivering Tabletop Exercises}},
    booktitle = {29th Conference on Innovation and Technology in Computer Science Education},
    series    = {ITiCSE '24},
    publisher = {ACM},
    address   = {New York, NY, USA},
    year      = {2024},
    pages     = {220--226},
    numpages  = {7},
    isbn      = {979-8-4007-0600-4},
    url       = {https://doi.org/10.1145/3649217.3653642},
    doi       = {10.1145/3649217.3653642},
}

@InProceedings{Kavrestad2025,
author="K{\"a}vrestad, Joakim
and Johansson, Sonny
and Bergstr{\"o}m, Erik",
editor="Oliva, Gabriele
and Panzieri, Stefano
and H{\"a}mmerli, Bernhard
and Pascucci, Federica
and Faramondi, Luca",
title="{Using Tabletop Exercises to Raise Cybersecurity Awareness of Decision-Makers}",
booktitle="Critical Information Infrastructures Security",
year="2025",
publisher="Springer Nature Switzerland",
address="Cham",
pages="231--248",
url={https://doi.org/10.1007/978-3-031-84260-3_14},
isbn="978-3-031-84260-3"
}

@article{chernikova2020simulation,
  title={{Simulation-Based Learning in Higher Education: A Meta-Analysis}},
  author={Chernikova, Olga and Heitzmann, Nicole and Stadler, Matthias and Holzberger, Doris and Seidel, Tina and Fischer, Frank},
  journal={Review of educational research},
  volume={90},
  number={4},
  pages={499--541},
  year={2020},
  publisher={Sage Publications Sage CA: Los Angeles, CA},
  url={https://doi.org/10.3102/0034654320933544}
}

\end{document}